\documentclass{emulateapj}

\usepackage{graphicx}
\usepackage{epstopdf}

\newcommand{\lx}{erg~s$^{-1}$}
\newcommand{\nh}{cm$^{-2}$}
\newcommand{\chandra}{{\it Chandra}}

\newcommand{\msun}{$M_{\odot}$}

\shorttitle{Blue Supergiant XRBs in IC 10}
\shortauthors{Laycock et al.}

\begin{document}


\title{Blue Supergiant X-ray Binaries in the Nearby Dwarf Galaxy IC 10 }


\author{Silas G. T. Laycock \altaffilmark{1}, Dimitris M. Christodoulou \altaffilmark{1}, Benjamin F. Williams \altaffilmark{2}, Breanna Binder \altaffilmark{3}, and Andrea Prestwich\altaffilmark{4}}


\altaffiltext{1}{Lowell Center for Space Science and Technology, University of Massachusetts Lowell, 600 Suffolk Street, Lowell, MA 01854, USA. Email: silas\_laycock@uml.edu, dimitris\_christodoulou@uml.edu}
\altaffiltext{2}{University of Washington, Department of Astronomy, Box
351580, Seattle, WA 98195, USA. Email: ben@astro.washington.edu, bbinder@astro.washington.edu}
\altaffiltext{3}{Department of Physics and Astronomy, California State Polytechnic University, 3801 West Temple Ave, Pomona, CA 91768}
\altaffiltext{4}{Harvard-Smithsonian Center for Astrophysics, 60 Garden Street, Cambridge, MA, 02138, USA. Email: aprestwich@cfa.harvard.edu}



\begin{abstract} 

In young starburst galaxies, the X-ray population is expected to be dominated by the relics of the most massive and short-lived stars, black-hole and neutron-star high mass X-ray binaries (XRBs). In the closest such galaxy, IC 10, we have made a multi-wavelength census of these objects.  Employing a novel statistical correlation technique, we have matched our list of 110 X-ray point sources, derived from a decade of Chandra observations, against published photometric data.  We report an 8$\sigma$ correlation between the celestial coordinates of the two catalogs, with 42 X-ray sources having an optical counterpart. Applying an optical color-magnitude selection to isolate blue supergiant (SG) stars in IC 10, we find 16 matches. Both cases show a statistically significant overabundance versus the expectation value for chance alignments. The blue objects also exhibit systematically higher $f_x/f_v$ ratios than other stars in the same magnitude range.  Blue SG-XRBs include a major class of progenitors of double-degenerate binaries, hence their numbers are an important factor in modeling the rate of gravitational wave sources. We suggest that the anomalous features of the IC 10 stellar population are explained if the age of the IC 10 starburst is close to the time of the peak of interaction for massive binaries. 
\end {abstract}


\section{Introduction and Motivation}

The dwarf starburst galaxy IC 10 is remarkable for its young stellar population \citep{Massey2007}, high density of Wolf-Rayet (WR) stars \citep{Crowther2003}, and massive stars in general \citep{Massey2002}. It hosts the WR+BH (black hole) binary IC10 X-1 \citep{Prestwich2007}, the blue-supergiant (BSG) transient X-ray binary IC 10 X-2 \citep{Laycock2014}, and an excess of X-ray point sources concentrated within the optical outline of the galaxy \citep{Wang2005}.  We have identified more than a dozen other transient and variable objects among 110 point sources detected in a series of 10 \chandra~ observations spanning 2003-10 \citep{Laycock2016}. The ongoing starburst has produced a rich population of exotic compact-object binaries. Given that the age of the starburst is only $\sim$6~My, these must involve progenitors of extreme masses and short lifetimes. Identifying the nature of each individual source is a slow and difficult process, but ultimately necessary if we are to obtain a complete census of X-ray binaries (XRBs) for comparison with other galaxies, for example the Magellanic Clouds \citep{Coe2015,Antoniou2010} and the local volume survey \citep{Binder2015a}. Even to obtain the X-ray luminosity function (XLF) requires these identifications, otherwise the $logN-logS$ relation will be heavily biased by foreground Galactic interlopers. We have adopted a multi-wavelength approach in order to narrow down the probable nature of the X-ray population in IC 10. 

Compact and dwarf galaxies present the opportunity for ``whole galaxy" population studies, and can thus probe a wide parameter space in age, mass, and metallicity.  The production rate of high mass X-ray binaries (HMXBs) is linked to the star formation rate (SFR) by a scaling law \citep{GGS}, the universality of which raises interesting questions.  Using a large sample of blue compact galaxies, \cite{Brorby2014} revealed a significantly higher abundance of HMXBs for a given SFR than in spiral galaxies. Additionally, the slope of the XLF is found to be steeper in the blue compact sample, in the sense that the XRB population is skewed to higher $L_X$ values in addition to being more numerous on a per-SFR unit basis.  \cite{Prestwich2013} found from their analysis of a large sample of galaxies that ultraluminous X-ray sources (ULXs) ($L_X > 10^{39}$\lx)  are preferentially formed at low metalicities.  

The inference preferred by many authors is that low metallicities drive higher compact-object masses,  as for example in the direct-collapse scenario of \cite{Mapelli2010}. Indeed, stellar evolution calculations also predict such a relationship \citep{Belczynski2010}. A counterargument however would be that the stellar companions in these systems are of a different nature (or have different binary parameters) such that systematically higher mass-accretion rates are delivered.  In support of this idea, we point to recent studies of bright extragalactic XRBs and ULXs such as M82 X-2 \citep{Bachetti2014}, NGC 300 X-1\citep{Binder2015b}, and IC 10 X-1 (\citealt{Laycock2015,Steiner2016}). These cases are actively recasting the paradigm of ULXs as extreme binaries hosting black holes of $>100$ \msun, as instead representing the upper end of continuous XLFs for neutron-star (NS) and BH binaries; perhaps the same skewed XLF discovered by \cite{Brorby2014}. 

Testing these scenarios requires nearby galaxies where the stellar companions can be resolved and optical spectra can be obtained.  IC 10 is the closest to being an ideal laboratory for such a study since its O and B stars are within range of ground-based telescopes, yet its angular size ($<$8 arcmin) still fits within a single \chandra ~ACIS-S3 field.   

IC 10 is a unique galaxy and reconciling the strange features of its stellar population and structure will provide further insight into XRB formation. Researchers using different SFR indicators have reported wildly differing results for IC 10 \citep{Chomiuk2011}. The ``brightest radio SNR" method yields the highest rate which ties in with the abundance of massive stars.   \cite{Leroy2006} suggests that the starburst has not yet peaked, and since its WR stars each represent a future type II supernova, IC 10 has produced only a small fraction of its eventual supernova remnants and compact objects. The work of \cite{Eldridge2013} on the evolution of massive binaries points to interaction as an underestimated driver of core-collapse supernovae (SNe). In essence, mass-transfer between the two stars narrows (or even reverses) their mass-ratio causing the initially less massive star to undergo an accelerated evolution. Although many scenarios are possible depending on the initial mass ratio and separation, the main result that emerges is a greatly enhanced population of blue and yellow supergiant binaries at the pre- and post-core-collapse stages. Enhanced tidal stripping of the hydrogen envelopes also leads to elevated production of WR stars.  Interestingly,  the \cite{Massey2007} HR diagram for IC 10 shows a large population of blue and yellow supergiant stars, in addition to the abundance of WR stars already noted. 

 HMXBs containing blue supergiants (incl. WR stars) are expected to be a leading channel for production of double-degenerate binaries (DDs) of the BH+BH, BH+NS, NS+NS species \cite{Bulik2011,Bogomazov2014}. If a massive binary system survives the first supernova and the remaining star is sufficiently massive, it too may undergo supernova, with some probability of each star ending as either BH or NS, while remaining gravitationally bound. A wide range of evolutionary paths have been studied, some involving mass transfer, common envelope, or other binary interaction. As in the \cite{Mandel2016} mechanism for creating double black holes by homogeneous chemical evolution driven by rotational mixing of the stars' interiors.  Secure DD discoveries include the  Hulse-Taylor pulsar (NS+NS), the WD+NS pulsar of \cite{Antoniadis2013}, and some 10s of binary WDs \cite{Gianninas2015}. The recent advent of gravitational-wave astronomy has revolutionized this field, piquing interest in the production channels for DDs, and event rates for their mergers, which are now observable as GW signals \cite{Abbott2016}.  Production and merger rates for DDs have been modeled by several groups (e.g. \citealt{Dominik2015,Abbott2016,Eldridge2016,Belczynski2016}) considering a wide variety of evolutionary paths, some highly modified by binary interaction. IC 10 can be seen as a laboratory for obtaining improved inputs to these models.

In this paper, we focus on the statistical correlation between X-ray and optical catalogs and on the resulting photometric properties of the sample of positional counterparts. We begin by matching source lists, combining the optical and X-ray photometry, then make a series of selections on the observable source properties. Finally, we assess the correlations that emerge between these subsets. \\

\section{Optical Counterparts and Color-Magnitude Selection for Blue Supergiants}
\label{sect:optical}
\begin{figure}
\begin{center}
\includegraphics[width=11.4cm,trim={6cm 4.5cm 1cm 4cm}]{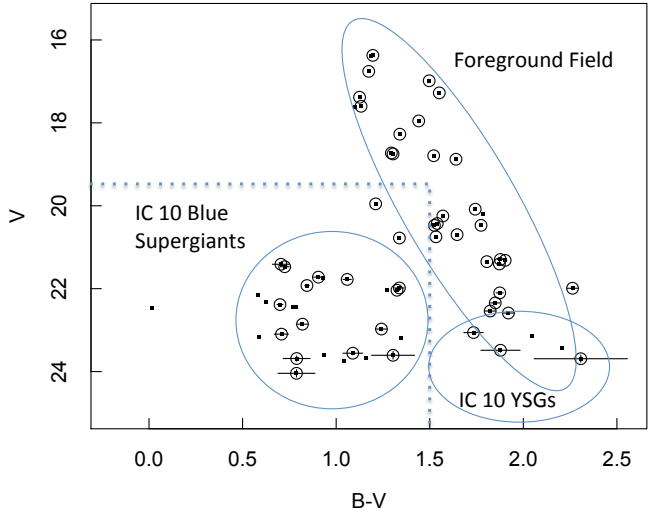}
\caption{Color-Magnitude diagram for positional counterparts to Chandra X-ray sources in the IC 10 field. Ovals indicate the locii of BSGs in IC 10 and the foreground Galactic population which lies atop the yellow SG track for IC 10 \citep[comparison with][Figure 14]{Massey2007}. A few objects have multiple counterparts; then the one at the smallest offset is retained and the others are indicated by a smaller dot symbol. Dotted lines show the selection B-V$<$1.5, V$>$19.5.}
\label{fig:colormag}
\end{center}
\end{figure}

Following the procedures detailed in \cite{Laycock2014}, we matched our Chandra point-source catalog against the \cite{Massey2007} photometric catalog which has a limiting magnitude of V$\approx$24. A total of 42 X-ray sources (40\%) have a positional counterpart within their 95\% confidence radius (in quadrature with a 1$^{\prime\prime}$ tolerance to account for systematic error). The remaining 69 sources have no counterpart to down to V=23.8 (the magnitude of the faintest counterpart found). In Figure~\ref{fig:colormag}, we show the color-magnitude diagram (CMD) for these candidate counterparts.  At the distance and reddening  of IC 10 ($\mu$ = 24 , $A_V$= 3, $E_{B-V}$= 0.85), the main sequence is not visible much beyond B0V in ground-based images under normal seeing conditions.  BSGs are among the most luminous stars and will be above the crowing limit, nonetheless this is a limitation for our HMXB study since the literature points to most HMXB counterparts being hotter than B3 with the spectral type distribution peaking at B0 \citep{Antoniou2009}. Therefore we estimate the catalog to be about 50\% complete for HMXB counterparts. The line of sight to IC 10 passes though the outer part of the Galactic Plane, thus  foreground stars lie in the field. Fortunately, the reddening vector is such that the main sequence defined by the Galactic stars is well separated in color-magnitude space from the IC10 BSG branch.  This fact was analyzed in detail by \cite{Massey2007} and we follow their Figure~14 (of which our Figure~\ref{fig:colormag} is a subset of the same data) to annotate the CMD highlighting the locus of IC 10 BSGs and the foreground sequence.

\begin{figure}
\begin{center}
\includegraphics[angle=0,width=9.1cm,trim={0 0.5cm 0 2cm}]{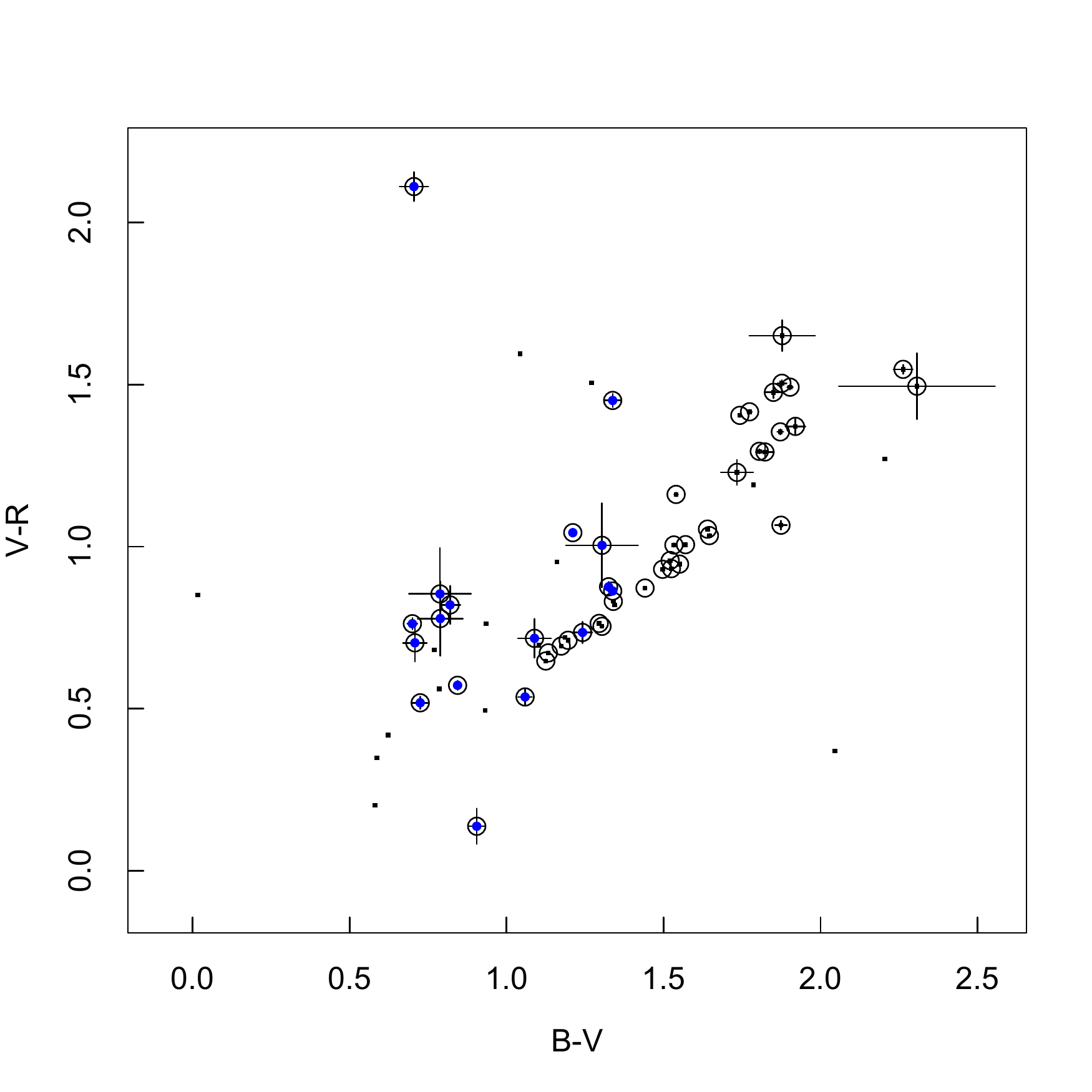}
\caption{Color-color diagram for positional optical counterparts. Objects selected as BSG candidates in Figure~\ref{fig:colormag} are highlighted as filled blue symbols.  As a few objects have 2 counterparts, the one at the smallest offset is retained and the others are indicated by a smaller dot symbol. Several of the BSG stars are red in R-I, indicative of an infrared excess that could be related to circumstellar material (outflows, disks, etc.).}
\label{fig:colorcolor}
\end{center}
\end{figure}

Having selected our sample of BSG-XRB candidates, we examined their photometric properties in Figure~\ref{fig:colorcolor} which is the color-color diagram (B-V, V-R). All counterparts are plotted and the BSG subset is highlighted. We see the expected linear sequence defined by the majority of the optical counterparts. Interestingly, many of the blue objects in B-V are significantly redder in V-R compared to the blackbody locus. This effect cannot be interstellar absorption because the stars are already bluer in B-V than most of the foreground population and absorption is stronger at shorter wavelengths. Similarly, since the BSGs are the bluest objects in B-V, they should also be the bluest in V-R if their continua follow an approximate blackbody spectral energy distribution.   
This red excess is possibly related to outflows from the companions and circumstellar material intrinsic to the binaries.  Alternatively AGN could be present in the sample and occupy similar regions of the CMD. Spectroscopic and radio followup are needed to explore these possibliities.

\section{Peak-Up Test for Positional Correlation}
Optical counterpart identification in multi-wavelength surveys is a problem  encountered frequently in astronomy. Often researchers assume that positional alignment implies physical association, especially when the objects being searched for are rare and both catalogs have good positional accuracy. This is a dangerous assumption and a needless one, since a quantitative solution to the problem does exist:  the peak-up test \citep{Laycock2005, Zhao2005} can be used to deal with the case of \chandra~ observations of crowded fields where chance alignment is non-negligible and each X-ray source has a different positional uncertainty. 

\begin{figure}
\begin{center}
\includegraphics[angle=0,width=8.5cm,trim={0 2.5cm 0 3cm}]{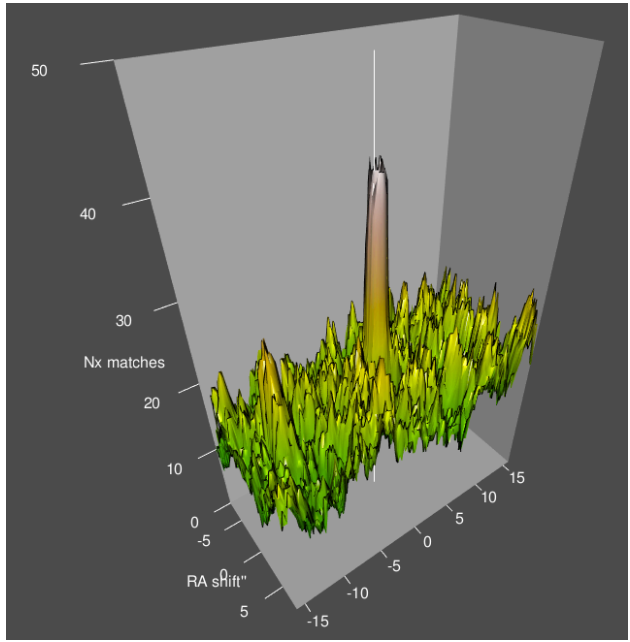}
\caption{Peak-up test surface plot showing the cross-correlaton between our \chandra  ~X-ray catalog and the ~\cite{Massey2007} optical catalog.  The surface denotes the number of sources with an optical counterpart, generated by matching the two catalogs over a grid of positional offsets. A sharp peak appears at the correct alignment, demonstrating that the optical counterparts are real. The variations in the surface, map the random fluctuations in the number of chance alignments. The distribution of the surface heights yields $\sigma$ and hence the correlation significance. For {\it all stars}, the correlation peak is 8.4$\sigma$; for the BSGs alone, it is 3.24$\sigma$. }
\label{fig:peakupsurface}
\end{center}
\end{figure}

In the peak-up test, the two input catalogs are repeatedly matched against each other, introducing a small offset in the coordinate system at each iteration, and recording the list of matching objects at each offset. The offsets are arranged on a regular grid whose spacing is finer than the positional accuracy of the input catalogs and extends to several times the radius of the largest error circles. In this way, the distribution of chance alignments is obtained free of any real ones.  The output table can then be filtered according to any desired properties of the input catalogs. Our code uses tools from the unix relational database package {\it starbase}\footnote{http://hopper.si.edu/wiki/mmti/Starbase} for the matching and filtering and $R$\footnote{https://www.r-project.org} for analysis and visualization of the resulting peak-up table. If an excess of real counterparts are present above the median number expected by chance ($\bar{n}$), a peak of height $n_x$ is seen in the 2D map, as in Figure~\ref{fig:peakupsurface}, hence the name ``peak-up" test. The significance of the peak is assessed by computing the standard deviation ($\sigma$) of the distribution of chance alignments after excising the region within the maximum matching radius of the zero-offset position. The significance of a positive correlation peak can be expressed as $S=(n_x - \bar{n})/\sigma$. For our Chandra catalog versus \cite{Massey2007}, using a matching radius for each source of $r=\sqrt{ r_{95}^2 + 1^2}$ (arcsec), we obtained $n_x$=42,  $\bar{n}$=16, $\sigma$=3.1, and $S$=8.4 

\begin{figure}
\begin{center}
\includegraphics[angle=0,width=8cm]{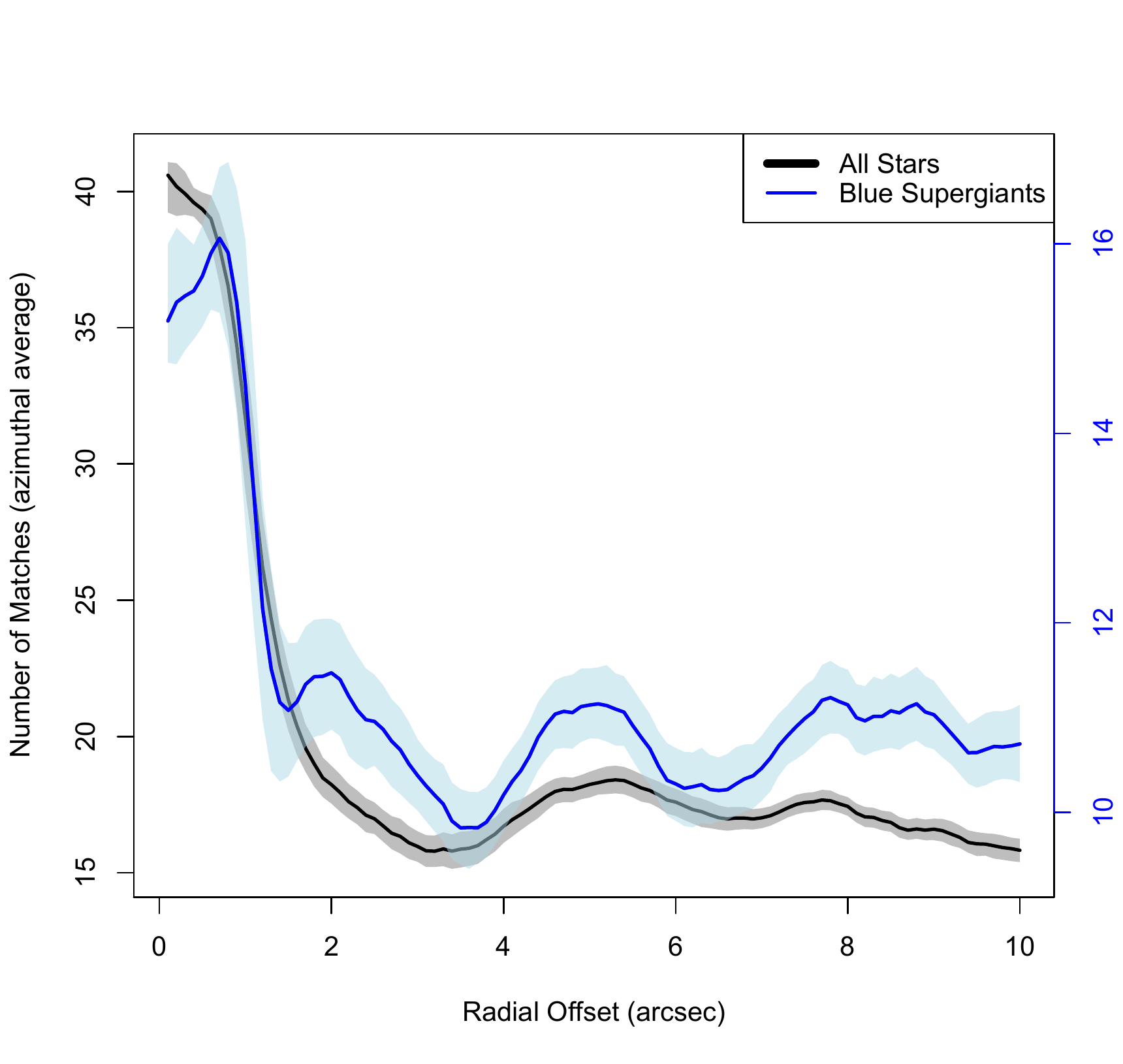}
\caption{Cross-correlaton between our \chandra ~catalog and the optical catalog of Massey 2007. In order to accumulate the plotted radial profile and the statistical distribution of chance alignments, the catalogs were repeatedly matched on a large grid of positional offsets as in Figure~\ref{fig:peakupsurface}.  
  Radial profiles are plotted for all stars  (black, left-hand scale) and the blue supergiant subset (blue, right-hand scale). 95\% confidence regions determined by bootstrap sampling with a kernel bandwidth of 0.25$^{\prime\prime}$  are shown as shaded areas. The narrowness of the correlation peak ($<2^{\prime\prime}$) indicates that crowding is only moderate in the magnitude range being probed, supported by the fact that more than half of the X-ray sources have no counterpart down to V=24.}
\label{fig:radialprofile}
\end{center}
\end{figure}

Since the candidate counterparts depicted in Figure~\ref{fig:colormag} fall into distinct regions of the CMD, identified by \cite{Massey2007} as BSGs and foreground main sequence stars, we applied a color-magnitude filter to the output of the peak-up test.  Adopting $B-V < 1.5$, and $V<19.5$ (see dotted lines in Figure~\ref{fig:colormag}, based on the photometric properties of the IC 10 population and the optical counterpart of IC 10 X-2, we obtain $n_x$=16,  $\bar{n}$=9, $\sigma$=2.16, and $S$=3.24. This result is shown graphically in Figure~\ref{fig:radialprofile}, where the surface defined by the peak-up test has been azimuthally averaged using a gaussian kernel (bandwidth=0.1 arcsec) to obtain the radial profile of the distribution. We can further assess the significance of the peak using the fluctuations of the resulting radial distribution. We find $\sigma_{r}$ = 0.97, $S_r$  = 24 for all counterparts, and $\sigma_{r}$ = 0.56, $S_r$  = 8 for BSGs. 

 Confidence regions for the peak-up test are constructed using bootstrap sampling in {\it R}. At each iteration we draw half of the offset positions in the peak-up table at random and compute the radial profile (as above). The process is repeated (10$^4$ trials) to accumulate the probability distribution, from which confidence regions are extracted using the {\it quantile} function. 
		
\section{Discussion}
\subsection{Evidence for blue supergiant X-ray binaries}
First we must establish that our sample of X-ray selected BSGs are accreting compact-object binaries since all BSGs are strong sources of X-rays \citep{Berghoefer1997}. Wind-origin X-rays are seen in both isolated and binary systems, however for the X-ray luminosities visible at the 660 kpc distance of IC 10, the only viable power-source is accretion onto a compact object. For B3 -- O5 spectral types, the intrinsic wind-origin $L_x = 10^{31} - 10^{33}$ \lx, which is below the $\gtrsim10^{35}$ \lx ~limit of the \chandra ~observations comprising our catalog.  

\begin{figure}
\begin{center}
\includegraphics[angle=0,width=9cm]{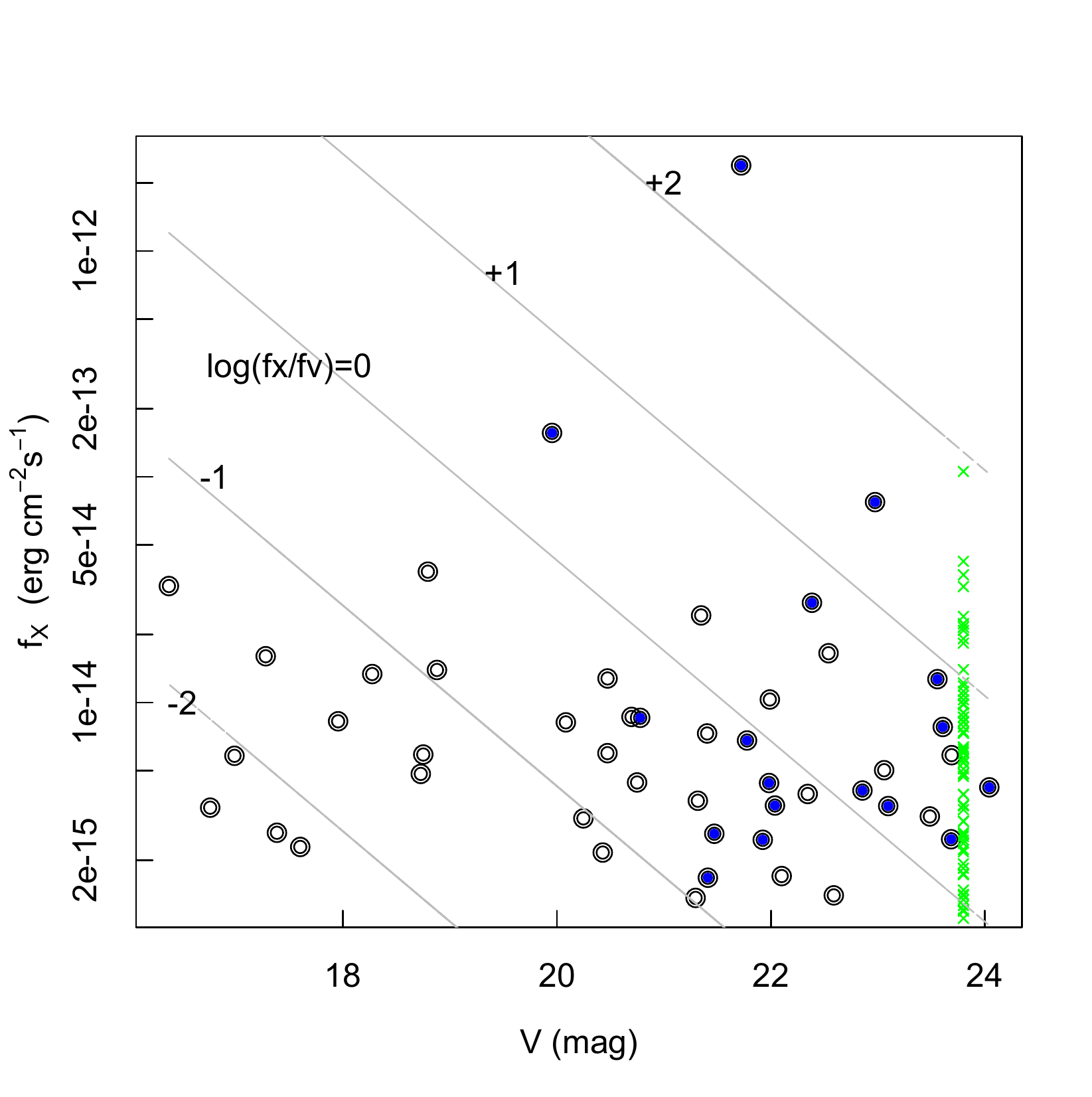}
\caption{X-ray flux versus V magnitude for X-ray selected stars in IC 10. BSGs are shown as filled symbols, other counterparts as open symbols. Orphan X-ray sources (without an optical counterpart) are plotted at the limiting magnitude of the catalog. Contours of constant X-ray to optical flux ratio $\log(f_x/f_v)$ are plotted for integer values from $-2$ to $+2$. Error bars for V values are $<$0.1~mag, while the flux ratios are the maximum values obtained from the brightest X-ray detections. }
\label{fig:fxfv}
\end{center}
\end{figure}

In order to put this comparison on a quantitative basis,  we computed the distance independent X-ray to Optical flux ratio  $\log(f_x/f_v) = \log(f_x)  + V/2.5 + 5.37$  ($f_x$ in the energy band: 0.3-8 keV) for all Chandra sources with counterparts in \cite{Massey2007}, as well as upper limits for sources with no counterpart in the catalog. Since the faintest optical counterpart is V=23.8, we adopted this limiting magnitude in the upper-limit calculation. Figure~\ref{fig:fxfv} shows the optical magnitude versus maximum broad-band (0.3-8 keV) flux from the 10 ACIS-S observations, assuming an absorbed power law ($n_H=5\times10^{21}$\nh, photon index $\Gamma=1.5$). Contours of equal  $f_x/f_v$ are shown and sources with BSG counterparts are flagged. Notably, the highest 5 $f_x/f_v$ values belong to BSGs, as are 7 of the top 9. For all stars in the same magnitude cut ($V>19.5$), the BSGs have systematically higher $f_x/f_v$ values which would not be expected due to chance. The MWW\footnote{Mann-Whitney-Wilcoxon test implemented in $R$.} test  shows a significant offset ($p=0.01$) for BSGs versus all counterparts. 

Five of the BSGs are X-ray variables as reported by \cite{Laycock2016} (source numbers 1, 8, 20, 29, 46). The $f_x/f_v$ points in Figure~\ref{fig:fxfv} are the maximum values obtained from the highest observed flux in each case. Variability ranges from a factor of a few to $\sim$100 (\#46), corresponding to changes in flux ratio $\Delta f_x/f_v$ of 0.3-2. Optical spectra have been published for source \#20 (WR+BH; IC10 X-1) and \#46 (IC 10 X-2). The variability properties of these sources were discussed in \cite{Laycock2016}. 
  
\cite{Massey2007b} made a search for LBVs in IC 10 using their broad-band photometry \citep{Massey2007} (the same catalog that we have employed in this paper) augmented with narrow-band filter observations (H$\alpha$, SII, OIII) in order to construct a multi-index based selection for intrinsically blue emission line stars.  Their resulting sample contained 96 candidates, {\it only one of which} has an X-ray counterpart (IC 10 X-2) in our \chandra ~catalog. This odd result is in line with those authors' puzzlement at the small number of WR stars in the emission-line sample. They explained the anomaly as an unintended consequence of applying selection criteria that were based on M31. In particular, Figure 3 of \cite{Massey2007b} shows that the cut on OIII emission designed to weed out HII regions had the effect of excluding 14/17 known WR stars in the field, along with a {\it majority} of the emission-line objects.  We also note that only one X-ray source (IC 10 X-1) has a position match in the WR catalog of \cite{Crowther2003,Clark2004}. A new emission-line catalog with relaxed criteria would be useful, however spectroscopic follow-up of all the optical counterparts will still be essential,  in view of possible contamination by AGN in the $f_x/f_v$ selected sample.

 \subsection{Estimating the underlying population}
If we assume that the $n_x=16$ BSG X-ray objects in Figure~\ref{fig:colorcolor} are all HMXBs, we can estimate their production rate
and the expected yield of precursor double-degenerate binaries $n_{PDD}$ as follows:

\begin{itemize}

\item[(a)]~If IC10 has produced $n_{PDD}$ precursor 
double-degenerate binaries over the past $T\approx 6$~Myr, then the duty cycle 
during which the LBVs are donating mass to their compact companions is $D=t_m/T$, 
where $t_m$ is the mean lifetime of the LBVs taken here to be 
$t_m=0.4$~Myr for a typical LBV mass of $M_2=30~M_\odot$ 
\citep{Maeder2001,Fragos2015,Massey2016}. In this case, we estimate that
$D=0.067$ and that $n_{PDD} = n_x/D = 240$ progenitors over the duration 
of the starburst.

\item[(b)]~We require a high mass ratio $q=M_2/M_1$,
where $M_1$ is the mass of the progenitor of the compact object and $M_1 > M_2$
so that the progenitor will indeed collapse first. \cite{Moe2013} found that
the close binary fraction of O/B stars with $P_{orb}<20$~days and $q>0.1$ 
in the low-metallicity Magellanic Clouds is $F_{cl}=0.22(M_1/10M_\odot)^{0.4}$,
from which we find using $M_1=35~M_\odot$ that $F_{cl}=0.36$. Furthermore,
\cite{Moe2013} find that the proportion $F_{tw}$ of ``twin" systems 
with $q>0.9$ is in the range of 0.06-0.16. We adopt a typical value of 0.1
for $q>0.9$, but we double this estimate to account for additional binaries 
whose mass ratio falls in the range of, e.g., $q=0.5-0.9$; so below we use 
an estimate of $F_{tw}=0.2$ for $q>0.5$.
Then we find that $n_{PDD}=n_x/(F_{cl}F_{tw}) = 222$ progenitors 
over the duration of the starburst.

\end{itemize}

Thus, it appears that the 16 observed BSG X-ray binaries in IC10 
indicate an underlying population of $\sim 200$ double-degenerate 
precursors and a production rate of $200/T\sim 30-40$~Myr$^{-1}$
over the past 6~Myr.  These estimates could be doubled if in case (b) one
chooses to double the lowest twin fraction observed in the Magellanic Clouds
\citep[0.06;][]{Moe2013} and uses $F_{tw}\approx 0.1$ instead of 0.2. The lifetime of the X-ray bright accreting phase is also subject to some uncertainty, but is in the range of 0.4~Myr from binary synthesis calculations \citep{Fragos2015}.   We caution that $n_x$=16 is an imperfect measurement of the number of BSG XRBs in IC 10 since the excess above chance $n_X - \bar{n} = 7$ and the true number lies between those values.  On the other hand, the optical catalog is about 50\% complete for HMXB secondaries in general (see ~\ref{sect:optical}), which correction goes in the opposite direction. Until spectroscopic identifications are completed, we treat $n_x$=16 as a working value.   
  
\subsection{Binarity and massive star evolution}  
Binarity affects the evolution of both stars in a close binary, with the outcome depending on the point in time at which the two stars interact. Stripping of the envelope of the primary occurs as it enters the giant phase, and the coincident accretion of this matter onto the secondary conspires to drive both stars toward the blue. This effect is supported by the fact that the progenitors of core-collapse SNe span a wider region of the HR diagram than do isolated massive stars \citep{Eldridge2013}. In the  case of SN 1993J, the progenitor star and the companion were resolved in pre- and post- explosion HST images \citep{Fox2014} showing just such a situation---the secondary having accreted the primary's hydrogen envelope prior to SN.  \cite{Eldridge2013} show that the excess of BSGs is generated along with a population of yellow supergiant binaries containing what are effectively subluminous, weak-lined WR analogs. The former binaries would naturally fall into the BSG region of the CMD (Figure~\ref{fig:colormag}), whereas the latter would reside into the area inconveniently occupied by the intersection of the IC 10 and foreground populations. The ratio of the two WR subtypes, WC and WN, has a strong metalicity dependence \cite{Eldridge2006};  the proportion of WN stars at SMC metallicity being an order of magnitude higher than at solar metallicity (data from M31). According to \cite{Kangas2016}, the spatial distribution of WN stars with masses $\gtrsim$20 $M_{\sun}$ in nearby galaxies matches that of SNe Ic. These are the most powerful type of core-collapse SNe and they are believed to be a privileged channel of BH production. On this basis, IC 10 is likely to be producing BH + massive star binaries and these would be among the X-ray selected objects in our sample. 

The subset that we have assumed to be foreground objects could also contain possible progenitor systems of Thorne-Zytkow objects (TZOs) in IC 10, long speculated to result from the merger of a NS with a companion via common envelope evolution or supernova kick.  With its overabundance of SG stars, IC 10 may be a good candidate for producing TZOs.   The strongest by far TZO candidate to date was found recently in the SMC by \cite{Levesque2014,Worley2016}, who point out that the key observational difference between a TZO and a normal RSG is the unique elemental abundance profile.  \cite{Tout2014} and \cite{Maccarone2016} provide alternative scenarios for producing intriguing TZO mimics, respectively in the form of super-AGB stars, and Galactic S-star + WD (or NS) binaries.  Yet more strong motivation for optical spectroscopic surveys in local-group galaxies.


\section{Conclusions}
This paper demonstrates that we are seeing a different XRB population in IC 10 than either the Milky Way or the Magellanic Clouds with the systems identified so far in IC 10 being dominated by BSG counterparts rather than Be stars. This contrast between galaxies is due to their different ages, metal contents, and star-formation histories.  Of the 16 X-ray selected BSG candidates reported here, five are X-ray variables ~\citep{Laycock2016}, of which two are known HMXBs. Spectroscopic followup and photometric monitoring are required to confirm the nature (spectral types, binary periods) of this new sample of extragalactic HMXBs in a very young starburst environment.  

Direct observations of the timescales for HMXB production have now been made in several nearby galaxies: the Magellanic Clouds \citep{Antoniou2010,Antoniou2016}, NGC 2403, and NGC 300 \citep{Williams2013}.  Most of these galaxies show a general agreement on a timescale of 40-70~Myr for the peak age of  HMXB production. A similar estimate comes from the population synthesis modeling of \cite{Fragos2015} for the progenitor of M82 X-2. However the Large Magellanic Cloud shows two younger subpopulations in the range of 6-25~Myr, whose confirmed SG-HMXBs peak at $\leq$10 Myr; whether this younger subpopulation contains Be stars is still unclear.  If little star formation has occurred in IC 10 prior to the current burst (which is not the case for the Large Magellanic Cloud with multiple subpopulations of ages ranging up to $>$100 Myr), then not enough time has elapsed for the Be phenomenon to develop. To test this hypothesis and show that no Be systems are present in IC 10, one needs to search substantially fainter (V$\simeq$26) in order to reach B3V, the latest spectral type known to occur in HMXBs.

Very young HMXBs may have substantially different counterparts, for example the supernova impostor SN 2010da has an age of $<$5 Myr, and it's quite unclear what the donor star is \citep{Lau2016}. Our recent Gemini spectroscopy appears that it may be in a transition state, as it exhibits both Be-SG and LBV behavior. This is one recent example of an extremely young HMXB with some sort of BSG companion \citep{Binder2016}. 

The identified mass-donor stars span several magnitudes in the visible spectrum. We can further consider the possibility that HMXBs are among the orphan X-ray sub-sample without any optical counterpart to $V\gtrsim24$. This limiting magnitude corresponds to $M_V \sim -3$ which would make Be stars later than B0V too faint to be well represented in the catalog of \cite{Massey2007}.

Our estimate of the production rate of BSG-HMXB is $\sim$200 over 6~Myr, the duration of the starburst, given estimates of LBV lifetime and the distribution of binary mass ratios. 
The production of binaries as a function of spectral class remains an active research topic. For early B type stars in particular, \cite{Moe2013} found a close binary fraction of $22 \pm 5 \%$ (for primary masses of $10~M_\odot$) and no evidence for metalicity dependence in either mass ratio or orbital period distribution. Other researchers have however identified strong metalicity effects in stellar rotation \citep{deMink2013} and mass-loss rates \citep{Vink2001}.   We hope that this paper will motivate researchers to carry out the observations and the model  calculations needed to improve upon the assumptions in our chain of reasoning. As an example, recent population synthesis calculations by \cite{Eldridge2016} show that mergers begin a minimum of 3~Myr after the onset of star formation and the merger rate rapidly rises to a peak of $\sim$0.1 merger per Myr per 10$^{6}$ $M_{\sun}$ at a time of about 10~Myr before gradually declining. The initial ``waiting period"  is essentially the time taken for stellar evolution to reach the point when mass-transfer begins in massive binaries. 

The emerging paradigm is that binary interaction plays a central role in the evolution of massive stars and the production of SNe. We conclude that the age of the IC 10 starburst is fortuitously such that massive binary interaction is approaching its peak.  The picture painted in \cite{Prestwich2015} and \cite{Brorby2016} advances the hope that the observable parameters can be organized into a ``fundamental plane'' in SFR, XLF, and metalicity. Populating such a plane requires deep multi-wavelength observations of star forming galaxies across age and metalicity space and  IC 10 is the youngest and most accessible case study. 

\cite{Eldridge2013} and collaborators have established the existence of a expanded region in the HR diagram populated by massive binaries.  The IC 10 red-giant branch intersects the foreground Galactic main sequence in the vicinity of $B-V=1.5$. It is therefore worth considering whether any of the X-ray selected optical counterparts in the overlapping region of the CMD are yellow supergiants and hence an additional component of the HMXB population. Optical spectroscopy can potentially test this hypothesis since radial velocity measurements will provide definitive answers.

\section{Acknowledgements}

This project was made possible by the support of SAO grant NAS8-03060 and the Physics Department of University of Massachusetts Lowell.


\end{document}